\documentstyle[aps,preprint]{revtex}
\begin{document}

\baselineskip=20pt

\thispagestyle{empty}

\vspace{20mm}

\begin{center}
\vspace{5mm}
{\Large \bf   Production and   Suppression of Charmonium in Nuclear Collisions
}

\vspace{2cm}

{\large \sf   Cong-Feng Qiao$^{a}$,
Xiao-Fei Zhang$^{a,b}$, and Wei-Qin Chao$^{a,b}$}\\[2mm]

{\small $^{a}$ CCAST (World Laboratory), Beijing, 100080, P.R. China }\\
{\small $^{b}$Institute of High Energy Physics, \\
Academia Sinica, P.O.Box 918-4, 100039, P.R.China}\\[20mm]

\begin{minipage}{100mm}

\centerline{\bf Abstract}
\vspace{5mm}

$J/\psi$ production cross section 
considering the contributions of both color-singlet 
and color-octet $c\bar c$ channels is calculated.
The result is  used to  study the suppression of $J/\psi$ 
in nuclear collisions. 
With  absorption cross sections 
for $(c \bar c)_8$ $\sigma_{abs}^8\simeq 11mb $
and $(c \bar c)_1$ $\sigma_{abs}^1\simeq 0mb $
the p-A and A-B data except for Pb-Pb can be explained.
Possible explanations of additional suppression 
in Pb-Pb  are discussed.

\end{minipage}
\end{center}

\vspace{20mm}
PACS number(s): 13.85.Ni,  25.75.Dw, 12.38.Mh

\newpage

\section{Introduction}

It was expected that a suppression of $J/\psi$ production in
relativistic heavy ion collisions can serve as a clear signature 
for the formation of a new matter phase Quark Gluon 
Plasma (QGP) \cite{Matsui}.
This suppression effect was observed by NA38 collaboration 
later\cite{NA38}. However, successive research pointed out 
that such suppression could also
exist in hadronic matter (HM), even though by a completely different
mechanism \cite{Capp}. The anomalous $J/\psi$ suppression  
recently reported by the NA50 collaboration\cite{na50} 
and there have been 
a number of attempts to explain  it,  such as the onset 
of deconfinement,
hadronic co-mover absorption and the energy 
loss model\cite{blqgp}\cite{comver}\cite{eloss}. 
To understand  the expermental data clearly, the 
formation and absorption mechanism of $J/\psi$ must be 
studied carefully. 

Quarkonium production has traditionally been calculated in the 
color singlet model. However, it has become clear now that the
 color singlet model fails to provide a theoretically and 
phenomenologically explanations  of all production processes
and the inclusion of color octet production channels removes large
discrepancies between experiment and the predictions of the color 
singlet model for the total production cross section.
In principle, the $J/\psi$ state is described in a Fock state
decomposition
\begin{eqnarray}
\label{Fock}
|J/\psi> &=& O(1)|c\bar c(^3S_1^{(1)})>+O(v)|c\bar c(^3P_J^{(8)}) g>
+O(v^2)|c\bar c(^1S_0^{(8)})g>+ \nonumber \\
&& O(v^2)|c\bar c(^3S_1^{(1,8)})gg>+O(v^2)|c\bar c(^3D_J^{(1,8)})gg>+...,
\end{eqnarray}
where $^{2S+1}L_J^{(1,8)}$ characterizes the quantum state of the $c\bar c$
with color-singlet or octet respectively. This expression is valid for
the non-relativistic  QCD (NRQCD) framework \cite{nrqcd}
and the coefficients of each
component depend on the relative three-velocity
$|\vec v|$ of the heavy quark, and under the limit of
$|\vec v|\rightarrow 0$, i.e. both $c$ and $\bar c$ remain relative rest,
eq.(\ref{Fock}) recovers the expression for color-singlet picture of
$J/\psi$ where $O(1)\equiv 1$.

The color-octet component $(c\bar c)_8$ seems to play  an
important role in interpreting the Collider Detector at Fermilab (CDF )
experimental data \cite{Braaten}, and 
further studies show that it may also have great influence on the quarkonium
production at other collider facilities \cite{a1,a2,a3}. The investigations
\cite{fixed} on quarkonium hadrproduction at fixed target energies find
that color-octet contribution to the cross section are very important,
and the inclusion of color-octet production channels removes large 
discrepancies between experiment and the predictions of the color-singlet
model for the total production cross section.
Therefore one can expect that the color-octet $(c\bar c)_8 $ would also 
manifest itself in heavy ion collisions\cite{satz8}. 
From the above discussion, one can know that 
the charmonium production can be devided into two steps.
The first step is the production of a $c \bar c $ pair.
The $c \bar c $ pairs can be either  $(c \bar c)_1$ 
or $(c \bar c)_8$. The $c\bar c $ pairs are produced 
perturbatively and almost instantaneously, with a formation time 
$\tau_f\simeq (2m_c)^{-1}\simeq 0.07 fm$ in the $c\bar c $ 
rest frame. 
The second step is the formation of a  physical states of  
$J/\psi$, that need much longer time.   People believed now that $J/\psi$ 
suppression in hadron matter can be consider
as pre-resonance absorption.
Satz first proposed to use the pre-resonance absorption model 
to explain nuclear collisions data \cite{satz8}. However, in their work 
the pre-resonance state of charmonia is only in color octet.
In this paper we give a complete leading order calculation of 
charmonium production  cross section for $(c \bar c)_1$ and
$(c \bar c)_8$
in corresponding energies. Then the results is 
used to study charmonium suppression in heavy ion 
collisions.
As it is known in the hadron-hadron collisions,
the $(c\bar c)_1 $ produced is   almost point like,  and    
the color-octet can interact with nucleons much more strongly than
the color-singlet $(c\bar c)_1$, so
the color-octet  would dissolve much faster into $D$ and 
$\bar D$ than $(c\bar c)_1$, i.e. $\sigma_{abs}^1\sim 0<<\sigma_{abs}^8$. 
Thus the absorption of $(c\bar c)_1 $
and $(c\bar c)_8 $ should be considered differently.
We find that in our model 
with the  absorption cross section  $\sigma_{abs}^1\simeq 0mb $
and $\sigma_{abs}^8\simeq 11mb $ the p-A and A-B data except for Pb-Pb 
can be explained. To explain  Pb-Pb data, additional
sources causing  nonlinear suppression in nuclear collisions
must be onset.

Our paper is devided into four parts.
In  section II, we describe the color-octet scenario borrowed from 
$p p$ collisions and calculate the charmonium production cross section
for $(c \bar c)_1$ and $(c \bar c)_8$.
In section III, the pre-resonance 
nuclear absorption model  considering  both 
$(c \bar c)_1$ and $(c \bar c)_8$ are described.
In the last section 
the result of our model is compared to the expermental data.
Some possible models for explaining 
the  anomalous $J/\psi$ suppression in Pb-Pb  collisions of NA50
experiment, such as the formation of QGP
are discussed.

\section{ Formulation}

According to the NRQCD factorization formalism, the inclusive production rate
of heavy quarkonium $H$ in parton level can be factorized as 
\begin{eqnarray}
\label{21}
\sigma (i j\rightarrow H ) = \sum\limits_{n}\hat{\sigma} (i j \rightarrow
Q\bar{Q}[n] ) <{\cal{O}}^H [n]>.
\end{eqnarray}
Here, $\hat{\sigma} (Q\bar{Q}[n] + X)$ describes the short distance production
of a $Q\bar Q$ pair in the color, spin and angular momentum state n, which
can be calculated perturbatively using Feynman diagram methods. $<{\cal{O}}^H
[n]>$, the vaccum expectation value of a four fermion operator in NRQCD 
\cite{nrqcd},
describes the nonperturbative transition of the $Q\bar{Q}$ pair hadronizating
into the quarkonium state $H$. 
The relative importance of the various contributions
of $n$ in eq.(\ref{21}) can be estimated by using 
NRQCD velocity scaling rules. 
An important feature of this equation is that $Q\bar Q$ pairs in a 
color-octet state are allowed to contribute to the production 
of a color singlet
quarkonium state $H$ via nonperturbative emission of soft gluons. Accordingly,
the production cross section for a quarkonium state $H$ in the hadron 
process
\begin{eqnarray}
\label{22}
A + B \rightarrow H + X
\end{eqnarray}
can be written as
\begin{eqnarray}
\label{23}
\sigma_H= \sum\limits_{i,j}\int^{1}_{0}dx_1 dx_2 f_{i/A}(x_1)
f_{j/B}(x_2){\sigma} (i j \rightarrow H ),
\end{eqnarray}
where the parton scattering cross section is convoluted with parton 
distribution functions $f_{i/A}$ and $f_{j/B}$, and the sum runs over 
all partons in the colliding hadrons.

At leading twist and at leading order in $\alpha_s$, 
the color-singlet $Q\bar Q$ production subproceses for $^{2S+1}L_J$ state
are 
\begin{eqnarray}
\label{24}
g  g \rightarrow ~^1S_0, ~^3P_{0,2},\\
g  g \rightarrow ~^3S_1+g, ~^3P_{1}+g,\\
g  q \rightarrow ~^3P_{1}+q,\\
q \bar q \rightarrow ~^3P_{1}+ g.
\end{eqnarray}
The corresponding formulae of the above precesses can be found in refs.
\cite{fixed} and \cite{report}. 
Because the radiative decays $\chi_{1,2}\rightarrow J/\psi + \gamma$ are 
known to have
a large branching ratios to $J/\psi$  and the feeddown of the
$\psi'$ to $J/\psi$ is also important, their 
contributions should be included in the calculation 
in reproducing the fixed target expermental data 
of prompt $J/\psi$ production.

Under the NRQCD factorization scheme, to calculate the quarkonium production
cross section, one use a double expansions: the perturbative expansion
of the short distance production amplitude in strong coupling constant 
$\alpha_s$ and the expansion of the nonperturbative 
long distance hadronization 
amplitude in typical velocity of heavy quark inside the heavy quarkonium.
At leading order in perturbative theory and up to next-to-leading order in the 
velocity expansion, the subprocesses for leading-twist $J/\psi$ production
through color-octet intermediate states are
\begin{eqnarray}
\label{25}
q  \bar q \rightarrow c \bar{c} [\b{8}, ^3S_1]\rightarrow J/\psi + X,\\
g  g \rightarrow c \bar{c} [\b{8}, ^1S_0]\rightarrow J/\psi + X,\\
g  g \rightarrow c \bar{c} [\b{8}, ^3P_J]\rightarrow J/\psi + X,\\
q  \bar{q }\rightarrow  c \bar{c} [\b{8}, ^3P_J]\rightarrow \chi_J + X 
\rightarrow J/\psi +\gamma + X.
\end{eqnarray}
Their cross sections are proportional to the NRQCD matrix elements
\begin{eqnarray}
\label{26}
<0|{\cal O}^{J/\psi}_8 (^3S_1)|0>\sim m_c^3 v^7,\\  
<0|{\cal O}^{J/\psi}_8 (^1S_0)|0>\sim m_c^3 v^7,\\  
<0|{\cal O}^{J/\psi}_8 (^3P_J)|0>\sim m_c^3 v^7,\\  
<0|{\cal O}^{\chi_J}_8 (^3S_1)|0>\sim m_c^3 v^5.  
\end{eqnarray}
It is obvious that the above matrix elements are higher order in $v^2$
compared to the leading color-singlet ones, but their corresponding 
short-distance processes are lower order in $\alpha_s$ than that in 
color-singlet processes, this makes the color-octet processes non-negligable
in reproducing the fixed target experiment data.

For $\psi'$ production the cross section does not receive contributions from
radiative decays of higher charmonium states, the $\sigma_{\psi'}$
differs from the direct $J/\psi$ production cross section 
$\sigma(J/\psi)_{dir}$ only in the  replacement of $\psi'$ matrix elements
 by $J/\psi$ matrix elements.

With the understanding of quarkoium production, 
in the calculation we can assume that the color-singlet and octet
$c\bar c$ are produced at early stage of
heavy ion collisions and later evolve
according to the environment. 
 Here we assume that  the
$c\bar c$ are produced at the early stage of the ion collisions through hard
N-N collisions
when the phase transition has not yet occurred, and then
the system would either remains in HM or a new phase QGP is formed at later
time. Thus we only need to deal with the evolution of $(c\bar c)$, while 
the production of $c\bar c$ by the deconfined quarks or gluons
in QGP is ignored.

\section{Absorption in  hadronic matter}
 
The conventional probability for a $J/\psi$ produced in a p-A collision
is given by:
\begin{eqnarray}
\label{abs}
R_A&=&{1\over A}{\sigma_{pA}\over \sigma_{pp}}\nonumber\\
&=&\int d^2b dz \rho_A(b, z)
exp\biggl\{-(A-1)\int_z^\infty d z'\rho_A(b,z')\sigma_{abs}\biggr\}\nonumber\\
&=&exp(-L_A \rho_0\sigma_{abs})
\end{eqnarray}
where $\sigma_{pp}$ and $\sigma_{pA}$ are the $J/\psi$ production 
cross section in proton-proton 
 collisions and proton-nucleus collisions, respectively,
$\rho_A$ is the nuclear density distribution. $\sigma_{abs}$
is an absorption cross section. L is the  effective length of the 
$J/\psi$ trajectory. It can be derived as
\begin{eqnarray}
L&=&{3\over 4}{A-1\over A}r_0 A^{1/3}, \ \  \ \ \  \ for \ heavy\ nucleus\nonumber\\
&=&{1\over 2}{A-1\over A}r_0 A^{1/3}{r_0^2\over {r'}_0^2},
 \  \   \ for \ light \ nucleus,
\end{eqnarray}
where $\rho_0=0.14 fm^{-3}$ and $r_0=1.2 fm, r'_0=1.05fm $

As $c \bar c$ pairs are produced almost instantaneously and the  
formation of a  physical states 
$J/\psi$ need a much longer time, people now believe that $J/\psi$ 
suppression in hadronic  matter can be considered
as an absorption of pre-resonance $c\bar c$ pairs. As discussed 
in former section, there are both $(c \bar c)_1 $ and $(c \bar c)_8 $
pairs.The $(c\bar c)_1 $ produced is  almost point like. 
The color-octet can interact with gluons much more strongly than
the color-singlet $(c\bar c)_1$,
and therefore would dissolve much faster into $D$ and 
$\bar D$ than $(c\bar c)_1$. Thus their absorption cross section 
are different. Considering these facts,
we extend the Eq.(\ref{abs}) to
\begin{eqnarray}
\label{absos}
R_A&=&{1\over A}{\sigma_{pA}\over \sigma_{pp}}\nonumber\\
&=&f_1\int d^2b dz \rho_A(b, z)
exp\biggl\{-(A-1)\int_z^\infty d z'\rho_A(b,z')
\sigma^1_{abs}\biggr\}\nonumber\\
&+&f_8\int d^2b dz \rho_A(b, z)
exp\biggl\{-(A-1)\int_z^\infty dz'\rho_A(b,z')\sigma^8_{abs}\biggr\}
\end{eqnarray}
Where $f_1, f_8$ are relative fractions of $(c \bar c)_1$
and $(c \bar c)_8$. $\sigma^1_{abs}, \sigma^8_{abs} $
are the absorption cross sections for $(c\bar c)_1$-nucleon     
and $(c\bar c)_8$-nucleon correspondingly.     

\section{Results and Discussions}

Before embarking on the computation of cross sections
the parameters used in the computation should be fixed up.
The value of $<0|{\cal{O}}_8^{J/\psi}(^3S_1)|0>$ can be obtained
by fitting the theoretical predictions to the 
CDF Collaboration data at large $p_T$. The number of 
independent matrix elements can be reduced by using the spin symmetry
relations up to corrections of order $v^2$
\begin{eqnarray}
\label{41}
<0|{\cal O}^{\chi_J}_1 (^3P_J)|0> = (2J+1)<0|{\cal O}^{\chi_0}_1 (^3P_0)|0>,\\
<0|{\cal O}^{J/\psi}_8 (^3P_J)|0> = (2J+1)<0|{\cal O}^{J/\psi}_8 (^3P_0)|0>,\\
<0|{\cal O}^{\chi_J}_8 (^3S_1)|0> = (2J+1)<0|{\cal O}^{\chi_0}_1 (^3S_1)|0>.  
\end{eqnarray}
Therefore, the matrix elements $<0|{\cal O}^{H}_8 (^1S_0)|0>$ and
$<0|{\cal O}^{H}_8 (^1P_0)|0>$ enter fixed target production of $J/\psi$ and  
$\psi'$ in the combination
\begin{eqnarray}
\label{42}
\Delta_8(H)\equiv <0|{\cal O}^{H}_8 (^1S_0)|0> + \frac{7}{m_Q^2}
<0|{\cal O}^{H}_8 (^3P_0)|0>.
\end{eqnarray}
Up to corrections in $v^2$, the relevant color-singlet production matrix
elements are related to radial wave functions at the origin or their
derivatives, 
\begin{eqnarray}
\label{43}
<0|{\cal O}^{H}_1 (^3S_1)|0> = \frac{9}{2 \pi}|R(0)|^2,~~
<0|{\cal O}^{H}_1 (^3P_0)|0> = \frac{9}{2 \pi}|R'(0)|^2,
\end{eqnarray}
which can be determined from potential model or from quarkonium
leptonic decays. 
The values of these parameters, which we use, are\cite{fixed}
\begin{eqnarray}
\label{44}
<0|{\cal O}^{J/\psi}_1 (^3S_1)|0> = 1.16~GeV^3,~~
<0|{\cal O}^{J/\psi}_8 (^3S_1)|0> = 6.6\times 10^{-3}~GeV^3,\\
<0|{\cal O}^{\psi'}_1 (^3S_1)|0> = 0.76~GeV^3,~~
<0|{\cal O}^{\psi'}_8 (^3S_1)|0> = 4.6\times 10^{-3}~GeV^3,\\
<0|{\cal O}^{\chi_0}_1 (^3P_0)|0>/m_c^2 = 4.4\times 10^{-2}~GeV^3,~~
<0|{\cal O}^{\chi_0}_8 (^3S_1)|0> = 3.2\times 10^{-3}~GeV^3, \\
\Delta_8(J/\psi)= 3.0 \times 10^{-2}~GeV^3,~~
\Delta_8(\psi')= 5.2 \times 10^{-3}~GeV^3.
\end{eqnarray}
We use the Gl\"uck-Reya- Vogt (GRV) leading order (LO )
\cite{grv} parameterization for the parton distributions of the protons.
The c quark mass is fixed to be $m_c=1.5$ GeV and the strong coupling is 
evaluated at the scale $\mu = 2 m_c$, that is $\alpha_s \approx 0.26$.
The results of the intergrated cross sections for color singlet and 
color octet channels at several different energies are listed in Table I.

In the equation (\ref{absos}) there are two parameters $\sigma_{abs}^1$ 
and $\sigma_{abs}^8$, which is different from 
Satz's model\cite{satz8}.
As ${c\bar c}_1$ produced is almost point like, $\sigma_{abs}^1$ 
is very small. Thus we  take $\sigma_{abs}^1=0$ and the 
value of $\sigma_{abs}^8$ is considered as an open parameter and 
determined such as to get the best agreement with the data.
In Fig.1 we see that with $\sigma_{abs}^8=11mb$  we get quite 
good agreement with the data except for Pb-Pb data.    
So our pre-resonance absorption model of color singlet and color 
octet can explains both the proton-nucleus and nucleus-nucleus
data up to S-U system. To explain the non-linear  suppression 
of $J/\psi$ additional source of charmonium in nuclear collisions
must be included. 

The anomalous suppression has been interpreted 
as a hint of QGP\cite{blqgp}.  If the QGP is formed, the interaction 
region consists of a hot part where a $J/\psi$ can be dissociated.
Ref\cite{blqgp} modelled the effect
of quark-gluon plasma formation by assuming
that the $J/\psi$ produced in those region  is completely destroyed
whereever  the density of the participants exceeds 
a critical value. The survival probability then becomes
\begin{eqnarray}
\label{qgp}
&&R_{AB}\nonumber\\
&&=f_1\int d^2b d^2b_A dz_A d^2b_B dz_B \rho_A(b_A, z_A)
t({\bf b }-{\bf b_A}-{\bf b_B} )
\rho_B(b_B, z_B)\theta(n_c-n_p({\bf b},{\bf b_A},
{\bf b_B})) \nonumber\\
&&
exp\biggl\{-(A-1)\int_{z_A}^\infty d z'_A\rho_A(b_A,z_A')
\sigma^1_{abs}\biggr\}
exp\biggl\{-(B-1)\int^{z_B}_{-\infty} d z'_B\rho_B(b_B,z_B')
\sigma^1_{abs}\biggr\}\nonumber\\
&&+f_8\int d^2b d^2b_A dz_A 
t({\bf b }-{\bf b_A}-{\bf b_B} )\rho_A(b_A, z_A)
d^2b_B dz_B \rho_B(b_B, z_B)
\theta(n_c-n_p({\bf b},{\bf b_A},
{\bf b_B})) \nonumber\\
&&exp\biggl\{-(A-1)\int_{z_A}^\infty d z'_A\rho_A(b_A,z_A')
\sigma^8_{abs}\biggr\}
exp\biggl\{-(B-1)\int^{z_B}_{-\infty} d z'_B\rho_B(b_B,z_B')
\sigma^8_{abs}\biggr\},
\end{eqnarray}
where $t({\bf b }-{\bf b_a}-{\bf b_B} ) $ is called the thickness 
function, $n_p$ is the density of participants per unit transverse area.
$n_c$ is the  critical value. If $n_c$  is chosen to be the
highest value attained in S-U collisions, i.e., $3.3 fm^{-2}$
as suggested in  Ref\cite{blqgp}, 
only in 66\%
part of the interaction region the $J/\psi$  can 
survive\cite{blqgp}.
Using this result, in Fig.2  we give a rough analysis of 
our model with QGP 
and it is compared with the result of 
our nuclear absorption model. From Fig.2 we can see that 
using the mechanism of QGP, the anomalous suppression in 
Pb-Pb can be explained.

However there have  been  other alternative explanations
of the Pb-Pb data by NA50, such as collisions with secondary
hadrons (co-mover) and the energy loss model\cite{comver}\cite{eloss}.
To distinguish these mechanisms much  more work  should be 
done in this field.     
\vskip 5mm
\centerline{\Large{\bf Acknowledgments}}
\vskip 5mm
This work was supported in part by the National
Natural Science Foundation of China and the 
Hua Run Postdoctoral Science Foundation
of China.

\vspace{2cm}

\noindent
\centerline{\bf Figure Captions}

Fig.1, The $J/\psi$ survival probability obtained using 
Eq.(\ref{absos}) with $\sigma_{abs}^8= 11mb $ 
and  $\sigma_{abs}^1=0mb $
compared to the expermental data of p-A and A-A
collisions at different energies.\\

Fig.2, The $J/\psi$ survival probability obtained using 
our nuclear absorption model 
for p-A and A-A collisions at 
$E=200 GeV$
and the result of  
Pb-Pb $E=158GeV$ without and with  QGP formation in 
34\%   part of the interaction region 
compared to the expermental data.\\

\vskip 1cm

\centerline {\bf Table Caption}

Table I. The intergrated cross sections for color singlet and color octet processes.

\newpage
\begin{center}

Table I

\vskip 1cm

\doublerulesep 2.0pt
\begin{tabular}{cccc} \hline \hline
 & E=450 GeV  & E=200 GeV & E=158 GeV  \\ \hline
$\sigma_1$ & 49.64 nb & 24.56 nb &19.54 nb\\ \hline
$\sigma_8$ & 95.66 nb &  55.48 nb &45.94 nb\\ \hline \hline
\end{tabular}

\end{center}
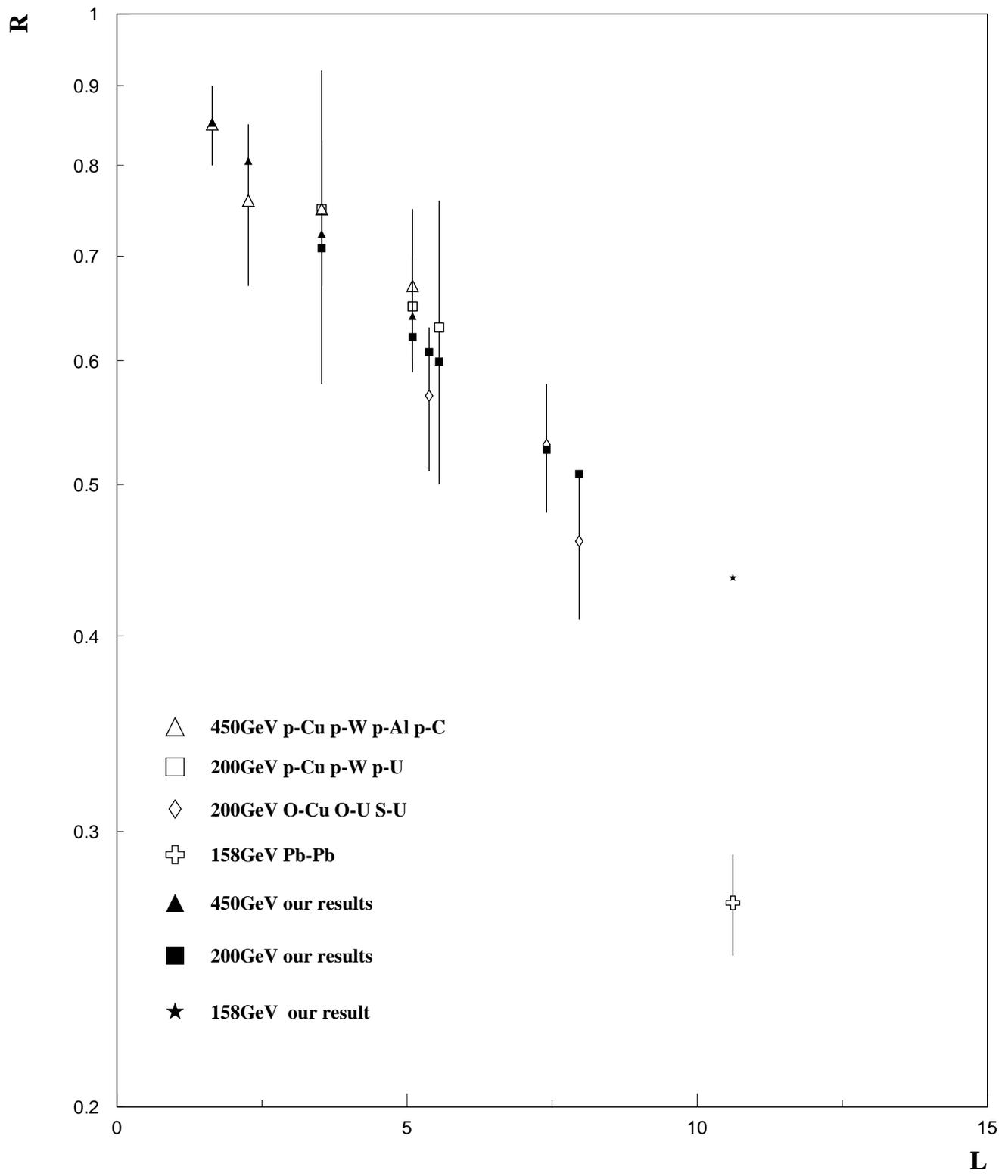
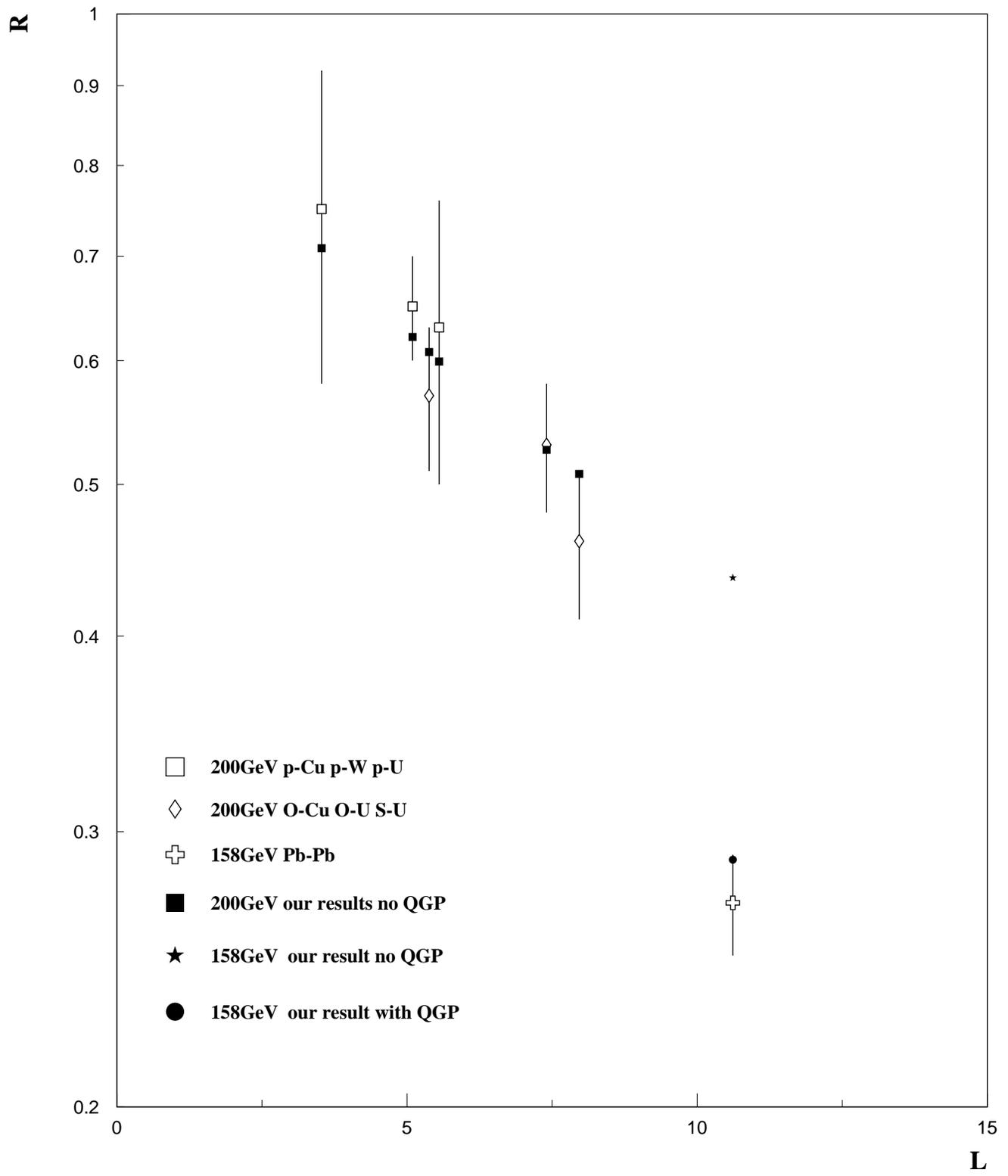
\end{document}